# Beyond method: The diatribe between Feyerabend and Popper over the foundations of quantum mechanics

Flavio Del Santo

Faculty of Physics, University of Vienna, Boltzmanngasse 5, 1090 Vienna (Austria) and
Institute for Quantum Optics and Quantum Information (IQOQI-Vienna), Austrian Academy of Sciences,
Vienna (Austria)

**Abstract**

Karl Popper and Paul Feyerabend have been among the most influential philosophers of science of the twentieth century. Extensive studies have been dedicated to the development of their controversial relationship, which saw Feyerabend turning from a student and supporter of Popper to one of his harshest critics. Yet, it is not as well known that the rift between Popper and Feyerabend generated mainly in the context of their studies on the foundation of quantum mechanics, which has been the main subject of their discussions for about two decades. This paper reconstructs in detail their diatribe over the foundations of quantum mechanics, emphasizing also the major role that their personal relationship played in their distancing.

1. **Introduction**

Both born and raised in Vienna in the aftermath of its intellectual golden age, Karl R. Popper (1902-1994) and Paul K. Feyerabend (1924-1994) have gone down in history as two of the most preeminent philosophers of science of the twentieth century. While Popper is famous for his solution to the problem of induction and demarcation based on falsifiability of scientific theories (see Popper 1934), Feyerabend –despite being a former pupil of Popper's– became renowned as one of the harshest and more distinguished critics of Popper's ideas, proposing instead methodological pluralism (called epistemological anarchism and later Dadaism; see Feyerabend 1975).

A vast literature has analyzed the intellectual relationship between these two preeminent philosophers of science (see, e.g., Preston 1997, Watkins 2000, Oberheim 2006, Collodel 2016) which led to identifying an "early Feyerabend", the staunch Popperian, and a "late Feyerabend", the sworn enemy of Popperism (see the comprehensive and excellent work Collodel 2016, and references therein). Yet, little attention –and only very recently– has been devoted to the dispute that involved Popper and Feyerabend over the foundation of quantum mechanics (FQM). However, as we shall see, FQM was the main subject of their interaction and it is in that context that the rift in their relationship developed (a first historiographical work in this direction was Del Santo 2019a). In fact, for almost two decades, from the early 1950s until the late 1960s, Feyerabend was foremost known as a philosopher of quantum physics, mostly building on positions of Popper and eventually turning against him. Despite a few earlier incidents, it is only at the end of the 1960s that Feyerabend firmly rejected Popper's viewpoint on quantum theory, fully embracing the Copenhagen interpretation, while dismissing Popper's criticism thereof as "surprisingly naive and uninformed" (Feyerabend 1968). This shift in Feyerabend's perspective is of great theoretical and historiographical interest because, as already hinted in (Del Santo 2019a), and further developed in (Kuby 2018) and (van Strien 2020), it is possible to claim that the recognition of the methodological validity of the development of Bohr's complementarity, on the one hand, and of Bohm's realist alternative interpretation on the other, provided the spark that led to the development of his methodological pluralism against Popper's philosophy of falsificationism.

However, by analyzing in detail the Popper-Feyerabend correspondence, now published in a commented collection (Collodel and Oberheim 2020), it is possible to identify a strong personal component of which the impact on the rift between the two should not be underestimated. As we shall see, it was hand in hand with their discussions about FQM that this personal development took place, and it is the aim of the present paper to reconstruct this side of the story. It should be remarked that the nature of the interaction between Popper and Feyerabend transcended that of a standard mentor-pupil relationship, reaching the level of a relationship that resembles that of an (authoritarian) father and a (rebellious) son. Thus, despite some reasonable explanations for Feyerabend's distancing from Popper's positions both in the philosophy of quantum theory and on matters of scientific method, this paper will show that Feyerabend's criticisms towards his former mentor stemmed from personal resentment too.[1]

## 2. Feyerabend and Popper, philosophers of quantum mechanics

When they first met in 1948 at the at that time recently established European Forum Alpbach, Popper had recently come back to Europe and was about to be appointed professor of Logic and Scientific Method at the London School of Economics (LSE). After a long period spent working on the philosophy of social science, he was just restarting dealing with topics of philosophy of science, and in particular of physics. Feyerabend, for his part, was a student of physics at the University of Vienna, with a broad interest in philosophy. Indeed, after remaining stuck with his physics thesis, and also thanks to the influence of Popper –with whom he had started a correspondence after the Alpbach Forum– he eventually graduated, in 1951, in the philosophy of science under Viktor Kraft, the last Vienna Circle member still in Vienna.

After his graduation, Feyerabend spent some time in summer 1952 in Denmark, where he listened to Niels Bohr's lectures and interacted with his collaborators, before being awarded a British Council scholarship to study FQM in London, under Popper's supervision, for the academic years 1952-53. Popper, in fact, had been giving contributions to quantum foundations since the early days of the theory, notably proposing a gedankenexperiment in 1934 aimed to show the untenability of Heisenberg's uncertainty principle, and that quantum particles can have both sharp position and momentum. His proposal turned out to be mistaken but it gathered the attention of, among others, Einstein and the Copenhagen school. After his return to Europe, Popper started again working on the foundations of physics, addressing the problem of indeterminism in classical and quantum mechanics (Popper 1951; see also Del Santo 2019a). Popper's position on the FQM can be summarized as advocating a realistic and indeterministic interpretation of quantum physics, openly antagonizing the widespread Copenhagen interpretation considered instrumental, positivistic and therefore anti-realistic (for a complete reconstruction of Popper's engagement in the FQM and his interaction with the physics community concerned with that topic, see Del Santo 2018; Del Santo 2019a; Del Santo 2019b; Del Santo 2020; and Del Santo & Freire 2021).

---

[1] For this reason, this paper would mainly use the correspondence between Popper and Feyerabend (Collodel and Oberheim 2020) as a source, to delve into their personal relationship.

Feyerabend, on the other hand, developed an early interest in the field of FQM with its mix of theoretical physics and philosophy, the two main subjects of his studies, most likely again thanks to his interaction with Popper. Since 1951, Feyerabend got interested in the so-called "von Neumann impossibility proof", a well-known result of John von Neumann that allegedly proved that no "hidden variable" deterministic completion (i.e. achieved by introducing additional variables besides those appearing in the state vector) can be consistent with the principles of quantum mechanics (von Neumann 1932). Feyerabend, together with Popper, started doubting its validity and their discussion on that matter continued for years.[2] In 1954, Feyerabend published his first paper on FQM, *Determinismus und Quantenmechanik* (Feyerabend 1954a), wherein he fully adhered to Popper's program by advocating indeterminism, even in classical physics, and for the first time casting doubt on von Neumann's impossibility proof in print.[3] The influence of Popper is there evident and explicitly acknowledged in the conclusions.

At the end of his scholarship, Feyerabend moved back to Vienna, while Popper –since at least as early as Summer 1953– strove for securing funds to bring back Feyerabend to London and promote him to his research assistant. Yet, although Feyerabend was initially enthusiastic, when Popper successfully got him a position, he suddenly changed his mind and turned down the offer. As a justification, he provided a conjunction of personal factors, as he wrote in a series of letters to Popper, finally stating: "I am very sorry that I still cannot give you any understandable account of the situation. None of the many reasons I have put before myself is really the decisive one. I only know that I would like to concentrate and improve my mental state." (Feyerabend to Popper 12.03.1954 N.11; in Collodel and Oberheim 2020). The main motivation can probably be sought in the fact that Feyerabend just got married and his wife was completing a PhD in Vienna (Feyerabend to Popper 03.02.1954 N. 6; in Collodel and Oberheim 2020). Moreover, Feyerabend adduced a further reason, which although it may sound like an excuse, it seems to have strongly conditioned his choices in accepting academic positions throughout his career also at later stages: his dream to become an opera singer. Indeed, he wrote to Popper: "I always wanted to be able to participate in music. […] I have already got an offer to play a solo part in the Konzerthaus [of Vienna] in June." (Feyerabend to Popper 03.02.1954 N. 6; in Collodel and Oberheim 2020).[4]

---

[2] The first mention of this topic in the Feyerabend-Popper correspondence is in a letter from July 1953 (Feyerabend to Popper 07.07.1953 N. 1; in Collodel and Oberheim 2020). Von Neumann proof has been the subject of long debate before and after Popper's and Feyerabend's criticisms. Eventually the generality of the proof was disproved by D. Bohm, J. S. Bell, S. Kochen and E. Specker among others. For a historical, critical reconstruction see (Dieks 2017).
[3] Popper had published (Popper 1951) and presented in a series of lectures in the US, also before Einstein and Bohr, the idea that fundamental indeterminism is present not only in quantum mechanics but in classical physics too (See Del Santo 2019a, and Del Santo & Freire 2021).
[4] Also in 1960, when Popper advised Feyerabend to apply to an open position in the UK and therefore leave his post in Berkeley, Feyerabend puts his passion in music before his career: "I think I shall stay in Berkeley because of my singing teacher. I shall not rest until I shall be able to enter the concert stage. (Feyerabend to Popper 03.04.1960 N. 3; in Collodel and Oberheim 2020).

In order to stress how much physics was the main focus of Feyerabend's work at the beginning of his career, it is interesting to notice that roughly at the same time, Feyerabend considered also applying for positions in pure physics: "I may get a scholarship for Harvard or Princeton […] I would mainly deal with pure physics (I very often thought whether it would not be a good thing for me to return for some time to pure physics – without loosing contact with philosophy. For philosophy will always be the main thing for me)." (Feyerabend to Popper 08.10.1953 N. 6; in Collodel and Oberheim 2020).

Despite the aforementioned rejection, Feyerabend and Popper kept interacting a great deal as pupil and mentor, carrying out several projects together. First of all, Feyerabend was entrusted by Popper to retrieve a lost manuscript of his *Logik der Forschung* (Popper 1934) in Vienna, as well as of the German translation of Popper's renowned book *The open society and its enemies* (Popper 1945). Moreover, the support that Popper gave to Feyerabend for him to find a suitable academic position in the following years almost goes beyond reason. Feyerabend wrote an enormous amount of letters to Popper asking for any kind of advice and letters of recommendation for applying to many different institutions, and Popper always promptly did it. And not only it was Popper that persuaded Feyerabend to apply to a position at the University of Bristol (Feyerabend writes to him: "I think I must be very grateful to you for it was really the discussion with you in July which made me decide in favour of Bristol"; Feyerabend to Popper 12.10.1955 N.12; in Collodel and Oberheim 2020), but it was Popper's and Schrödinger's support that eventually won Feyerabend an appointment as lecturer in the philosophy of science there in 1955 (see Collodel 2016).[5]

The Bristol years have been particularly important for Feyerabend's formation as a philosopher of quantum theory. In October 1955, indeed, Feyerabend wrote to Popper: "[FQM] are now the main field of my studies" (Feyerabend to Popper 14.10.1955 N.30; in Collodel and Oberheim 2020). He gave a lecture course at the University of Bristol in early 1956 on "Philosophical problems of modern physics", fully devoted to the FQM, which is remarkable for two reasons: on the one hand this shows how much Feyerabend was still adherent to Popper's realist and anti-Copenhagen program. Indeed, in his lectures, Feyerabend made use of the works of David Bohm, Alfred Landé, Luis de Broglie, and Henry Margenau, which were the main allies of Popper among the physicists in his realist campaign against the Copenhagen interpretation (see Del Santo 2019a). Popper was in those years Feyerabend's main reference figure on matters of quantum foundations. Indeed, Feyerabend wrote to him before his lectures, hoping to get soon the opportunity to receive his feedback: "I must prepare the lectures very carefully and I am a bit afraid of the famous people. Anyway I hope that the MS of at least the first three lectures will be finished in December such that I can show it to you." (Feyerabend to Popper 14.10.1955 N.30; in Collodel and Oberheim 2020).

On the other hand, more in general, this lecture course is of great interest because it may be one of the first university courses after World War II devoted entirely to quantum foundations. In fact, this topic became very unpopular among physicists after the war (see Freire 2014), but Feyerabend reported that his lectures were to be attended by several physicists (such as Maurice Pryce). For these reasons, it is worth publishing here almost entirely the (tentative) table of contents of these lectures as Feyerabend attached it to a letter which he sent to Popper (Feyerabend to Popper 28.10.1955 N.29; in Collodel and Oberheim 2020):

1) The role of statistical laws in classical physics

2) Causality and Quantum-Mechanics

---

[5] Feyerabend had met Schrödinger at least once in 1953. He nevertheless asked again Popper to get in contact with him on his behalf, who however politely replied: "I shall of course write to Schrödinger if you want me to but it is infinitely preferable that you should write to him yourself. I know that he remembers you, and it is a general rule that people prefer to be approached directly by the person most immediately concerned, rather than through an intermediary." (Popper to Feyerabend 01.05.1955 N. 11; in Collodel and Oberheim 2020).

The Neumann-proof and the objections brought forward against this proof by Grete Hermann, D. Bohm, de Broglie.

3) Is Quantum-Mechanics complete?

Discussion of the paradox of Einstein-Podolsky-Rosen […] and several attempts at its solution.

Discussion of the case of the particle in the box […] (This is a new attack, by Einstein,

against quantum theory. The book contains also an answer by

Bohm + critical remarks by de Broglie and Landè [*sic*] (the latter

quoting you).

4) Quantum-theory of measurement

The views of Dirac, Kemble, Margenau, v. Neumann, Einstein,

Blochinzew.

5) The Interpretation of Quantum-Mechanics

(A) Particle-Interpretation: de Broglie's théorie de l'onde pilote

Bohm's interpretation […]

The theory of the double solution

(B) Wave-Interpretations: Madelungs hydrodynamical model

The views of Schroedinger […]

The $\psi$-function is taken to be real, the

sudden jumps made understandable by

a new interpretation of special relativity.

(C) Complementarity

(D) Reichenbach

(E) Three-valued logics (Weizsaecker; Reichenbach; Neumann-

<u>Birkhoff</u>)

6) <u>Quantum-Theory and the Reality of the external world</u>

(or Berkeley redivivus)

When the lectures finally started, Feyerabend enthusiastically reported to Popper, declaring that he completely embraced his point of view:

> I gave my first lecture to the physicists, I was very nervous (there were about 60 people), Prof. Pryce interrupted me several times, twice I gave in although I was right –but on the main point I convinced people. And this <u>main point</u> was that it does not follow from any presentation of the uncertainty-rel. that atomic particles cannot have sharp position + sharp velocity at the same time! For since I had left you your insistence on this point set my mind working with the result <u>that I have changed most of what I am going to lecture in favour of your point of view</u>. This I did about 3 days before the lecture[s] started and I found thinking along your lines most exciting + fruitful. (11.02.1956 N. 1; in Collodel and Oberheim 2020)

And he goes on accusing Bohr of conservativism: "Why do people believe into [*sic*] blurred electrons? I said in my lecture –and I think this to be true– that <u>the philosophy of the orthodox is the philosophy of the older quantum-mechanics (of the Bohr-Sommerfeld conditions)</u> and that Bohr never got rid of this philosophy." (Feyerabend to Popper 11.02.1956 N. 1; in Collodel and Oberheim 2020).

In 1956, Feyerabend published a note on von Neumann's impossibility proof that caused a real stir between him and Popper, for the latter accused Feyerabend of plagiarism (see section 3). This surely created a first main tension in their relationship but did not undermine it. Only a few months later, however, in January 1957, Feyerabend distanced from his own (and Popper's) result, getting convinced that "there is much more to the von Neumann's proof" than Popper originally thought. On this note, he wrote to Popper: "if I am right your and Joske [Agassi]'s criticism which I have adopted myself is wrong." (Feyerabend to Popper 21.01.1957 N. 2; in Collodel and Oberheim 2020). It is legitimate to think that this turn of events could have been triggered also by the personal quarrel that Feyerabend had with Popper on this topic. This reassessment of von Neumann's proof coincides also with Feyerabend's first open criticism of Popper's position and the acknowledgement of a certain value to the school of Copenhagen:

> […] any interpretation of quantum-mechanics which takes into consideration the Born-interpretation only –and this seems to be true of your [Popper's] as well as of Einstein's interpretation– is bound to lead to wrong conclusions. It is to be admitted that Bohr's and Heisenberg's presentation [*sic*] of the situation (which seems to be based upon an ontology of classical states) are very unclear and can be easily attacked. […] [O]ne cannot rest content with a purely statistical interpretation of Heisenberg's principle. (Fezerabend to Popper 28.01.1956 N. 3; in Collodel and Oberheim 2020)

But a real turning point in Feyerabend's ideas on FQM and the beginning of the real departure from Popper can be traced back to 1957. In April of that year, the *Ninth Symposium of the Colston Research Society* took place at the University of Bristol, turning out to be "the first major event after World War II" on the foundations of quantum physics (Kožnjak 2018). The symposium gathered a number of prominent physicists –including Bohm who had just become Professor at the University of Bristol– and distinguished philosophers of science, including of course Feyerabend who was even involved in its organization (on that conference, see the proceedings Körner 1957, and Kožnjak 2018). Popper, who was invited but could not attend, sent a paper that was read by Feyerabend, testifying that their relationship was still quite close. Around the Colston Symposium, it is possible to identify three main elements that may have contributed to Feyerabend's transition.

Firstly, (i) Feyerabend presented a talk on the quantum theory of measurement (published as Feyerabend 1957) wherein he advocated a solution to the quantum measurement problem –i.e. when, why, and under what circumstances a single measurement outcome obtains among the possible ones– that would not involve a (physical) collapse of the quantum wave function. Rather, he proposed a "for-all-practical-purposes" solution to the realization of a single outcome, namely that macroscopic superpositions are practically indistinguishable from decohered single outcomes. It is likely that when preparing this paper, Feyerabend engaged more thoroughly than ever before with the literature on the subject matter, and in particular with Bohr's writings. This may have led him to a reappraisal of Bohr's complementarity that, as also recently claimed in (Kuby 2018), he found quite similar to his own solution to the measurement problem. Indeed, shortly after the symposium he confesses to Popper:

> […] there is much more in the Copenhagen-interpretation (as it has been discussed by Bohr, not by the Bohrians) than I thought some time ago when I did not know it well enough. Do you know that Bohr was extremely critical of both Heisenberg's and von Neumann's book because of the account of the 'reduction of the wave-packet' given in those two books? What I tried in my paper on measurement was to make use of this criticism as well as of Bohr's suggestion to regard measurement as an irreversible process which is represented, in a schematic way, by the 'reduction of the wave packet'. (Feyerabend to Popper 21.07.1957 N. 11; in Collodel and Oberheim 2020).

And the following year Feyerabend published a paper fully devoted to Bohr's complementarity (Feyerabend 1958). Note, however, that his position towards Bohr remained ambiguous throughout the following few years, and he kept being critical of the Copenhagen interpretation in several occasions.

Moreover, (ii) one of the main topics of discussion at the conference was the, at that time, new alternative interpretation of quantum theory due to David Bohm (1952), which for the first time restored to a certain extent determinism and realism (see Kožnjak 2018; Freire 2019). Although Feyerabend was critical of Bohm's approach during the conference, he befriended Bohm and in the next few years the two established a close intellectual relationship (see van Strien 2020). Contrarily to what seems to be implied in (Oberheim 2006) and in (van Strien 2020), however, while Bohm's influence may have played a role in pushing Feyerabend towards methodological pluralism because he started judging both Bohr's and Bohm's mutually incompatible approaches justifiable on solid methodological ground, it is doubtful that Bohm had any impact on Feyerabend's shift of perspective on the FQM. In fact, Bohm also developed a strong relationship with Popper until the late 1960s, and they always remained very aligned with each other in their struggle for realism and against the Copenhagen interpretation (see Del Santo 2019a; Del Santo 2019b).

Finally, (iii) the Colston symposium was particularly important for Popper too, despite his absence, because it is there that he firstly presented *in absentia* his new "propensity interpretation", which became his main view on quantum physics for the rest of his life. This is both a new way to interpret probability calculus and at the same time a proposed interpretation of quantum theory. Propensities are meant to be objective probabilities, i.e. relational properties in an indeterministic world that quantify the natural tendency of a system to realize a certain outcome in a given situation (Popper 1957; Popper 1959). Feyerabend, who was entrusted with reading Popper's paper at the symposium, had the opportunity to study it and made no secret of being very dissatisfied by Popper's new position on QM. Before the conference he wrote to Popper:

> I must emphatically disagree about the part devoted to quantum-mechanics. […] Hence I must say that, although your paper is most important for probability as well as for the interpretation of the classical statistical disciplines, it does not contain any contribution towards the main problem of quantum-mechanics […]. (Feyerabend to Popper 27.03.1957 N. 4; in Collodel and Oberheim 2020)".

And again a few days later:

> However important this substitution may be for the understanding of probability and its use within the classical disciplines, it does not lead to any additional clarification in quantum-mechanics, especially it does not contribute anything towards the solution of the "quantum-mess" such as subjectivity and the like. […] Whether a physicist adopts now the frequency interpretation, or your interpretation, he would IN BOTH CASES add some further arguments in order to show that the single system cannot be in a well defined state. (Feyerabend to Popper 30.03.1957 N. 5; in Collodel and Oberheim 2020).

Feyerabend's dissatisfaction with Popper's propensity interpretation may have contributed further to the fact that he distanced himself even more from Popper on matters of FQM.

In 1958, again under Popper's recommendation (see Popper to Feyerabend 12.03.1958 N. 4; in Collodel and Oberheim 2020), Feyerabend was offered an appointment at the University of California in Berkeley, where he remained for the rest of his career (he retired in 1990). In that period, he kept working on some new ideas on quantum theory that radically drifted away from Popper's and that would eventually be collected in the almost-a-hundred-page long paper (Feyerabend 1962). In particular, in 1958, he drafted the paper *Observationally complete theories: some observations on quantum theory*, in which he adopted a manifest anti-realist approach, stating "that any attempt to maintain realism upon the microscopic level will have to change quantum theory".[6] It is thus not surprising that Popper gave extremely negative feedback to Feyerabend's paper:

> I am sorry to say that it shocked me. You are strangely illogical: your summary contradicts the contents of your own paper. […] I am almost convinced that your argument is woolly and invalid. But even if it is valid, there can be no doubt that you have completely misrepresented (and misunderstood) your own results. I am sorry to have to say all this. But it is no good suppressing criticism. My criticism is so simple and trivial that you can be quite sure that it is correct. (Popper to Feyerabend 05.05.1958 N. 6; in Collodel and Oberheim 2020)

Moreover, Feyerabend wrote to Popper that he had criticized, in yet another paper (which is roughly the final version of Feyerabend 1962), Popper's statistical interpretation of the uncertainty relations (Feyerabend to Popper 01.02.1961 N. 1; in Collodel and Oberheim 2020). Finally, in a post scriptum to a letter to Popper in 1961, Feyerabend confessed the real change of perspective he had towards Bohr, whose philosophy he had fully redeemed:

> I have now come to the conviction that Bohr's interpretation can be defended in a manner that does not at all bring in positivism and that it is a very reasonable interpretation. I also hear that Bohr himself violently opposes being called a positivist. I was quite surprised. One should not underestimate Bohr. (Feyerabend to Popper 29.04.1961 N. 6; in Collodel and Oberheim 2020)

The ambiguous position of Feyerabend, who was somehow trying to keep a foot in two worlds –the realists with Popper at the forefront on the one hand, and the supporters of Bohr and the Copenhagen interpretation on the other– was becoming evident to the community, and put Feyerabend in an uncomfortable position, as he wrote to Popper:

> I was supposed to participate in a very interesting project to reconstruct the history of the quantum theory from interviews […].[7] However I was dropped from the project as it was discovered that mentioning my name leads to violent reaction in Copenhagen. I am between two chairs, Landé has just dissociated himself from me on account of my being too kind to Bohr, and Copenhagen does not want me either. Philosophy of science is not a very easy thing, I must say. (Feyerabend to Popper 17.08.1961 N. 9; in Collodel and Oberheim 2020).

---

[6] This unpublished paper (reproduced in Collodel and Oberheim 2020) was supposed to be published in in the Proceedings of the XII International Congress of Philosophy, but it was withdrawn by Feyerabend due to Popper's criticism and published only a part of (Feyerabend 1962).

[7] Feyerabend is here talking about the famous project "Sources for History of Quantum Physics" that involved, among others, his colleagues at the University of California, Thomas S. Kuhn, and Paul Forman.

Despite all of this, in the first years he spent in California, Feyerabend retrieved the old relationship with his former teacher. He started again seeking Popper's guidance about his own career, he taught courses based on Popper's work, and he largely popularizes Popper's ideas on scientific method in the US, writing to Popper: "I am glad to see that everywhere they start realizing the importance of your ideas. […] [M]y course on scientific method would be entirely based on your ideas". (Feyerabend to Popper 02.03.1959 N. 3; in Collodel and Oberheim 2020); and: "I have just returned from a trip to southern California (St. Barbara, San Bernardino, Pomona, Univ. of Southern California, University of California) lecturing on quantum-theory (one lecture) and scientific method (four lectures, mainly an exposition, always enthusiastically received, of your ideas:" (Feyerabend to Popper 22.03.1959 N. 5; in Collodel and Oberheim 2020). Popper was also visiting professor in Berkeley for about five months in 1962, where he had regular interactions with Feyerabend.

A graver clash between Popper and Feyerabend was however yet to come. In 1967, Popper wrote *Quantum Mechanics without the Observer*, which has been rightly called Popper's "most influential essay on the topic [of FQM]" (Howard 2004). Therein, Popper attacked the Copenhagen school and in particular, Bohr and Heisenberg, stating that his essay "is an attempt to exorcise the ghost called 'consciousness' or 'the observer' from quantum mechanics" (Popper 1967).[8]

Feyerabend immediately reacted to such a criticism of Bohr's views, publishing the paper "On a Recent Critique of Complementarity" (Feyerabend 1968), wherein he states:

> The publication of Bunge's Quantum Theory and Reality and especially of Popper's contribution to it are taken as an occasion for the restatement of Bohr's position and for the refutation of some quite popular, but surprisingly naive and uninformed objections against it. Bohr's position is distinguished both from the position of Heisenberg and from the vulgarized versions which have become part of the so-called 'Copenhagen Interpretation' and whose inarticulateness has been a boon for all those critics who prefer easy victories to a rational debate" (Feyerabend 1968).[9]

Interestingly enough, also two of the physicists whose works Feyerabend had largely appreciated a decade before, so much so that he based some of his lectures in Bristol on their ideas, namely Margenau and Landé, were also directly attacked by Feyerabend. While Landé wrote to Popper that Feyerabend's paper "contains a number of quite intemperate (and entirely mistaken) attacks against you and Bunge, the main argument being that you 'did not understand Bohr'" (Landé to Popper; published in Del Santo 2019a), Margenau directly addressed Feyerabend with a rancorous letter:

> I am amazed that you should feel called to defend a physicist like Bohr, whose work is understood by all. Those of us who think little of his complementarity principle do not underestimate its original heuristic force, but discard it because recent developments in which he did not participate have now made it

---
[8] For a critical summary of this essay, see (Del Santo 2019a).
[9] Some of the physical content of this paper is analysed in (Del Santo 2019a).

> pointless. And your archaic examples and superficial verbalisations, far from changing that situation, greatly emphasize it. […] Your quotation […] accuses me of a stupidity which working physicists resent. […] Whether you accept my distinction between possessed and latent or dormant observable is unimportant; men whose judgement I respect have thought it significant. But to brush it off as fancy terminology without probing the substance and, worse, to indict my students along with me in this context is less than generous. It is indeed a gesture I have not previously encountered in my long career at Yale. (Margenau to Feyerabend on February 17th, 1969. Published in Del Santo 2019a).

Also Feyerabend himself informed Popper of his attack, showing such a naïve tone that it is difficult, in hindsight, not to interpret it as almost a mockery:

> Enclosed a note I have sent to BJPS volume [eventually published as Feyerabend 1968, and Feyerabend 1969] and which is critical of your article on the quantum theory in Bunge's volume. We have here a few students of physics who are thoroughly critical of the way in which quantum theory is being taught today and who just managed to introduce an official course, run by them, in which they want to explore the weaknesses of the orthodox interpretation. I gave them copies of your article and I never saw such an enthusiastic response. […] They also asked me for my opinion (they were all in my course on philosophy of science). As a reply I wrote the enclosed note, which I also sent to BJPS to be published as a <u>discussion note</u> there. On rereading it I find that on various occasions I have expressed myself rather harshly […] but I don't think this will do any harm. (Feyerabend to Popper 04.10.1967 N. 2; in Collodel and Oberheim 2020).

As a matter of fact, this is the last extant letter between the two, and more than likely it was indeed the last letter that Popper and Feyerabend ever exchanged.

### 3. The personal relationship and the alleged thefts of intellectual property

In the previous section it has been discussed the interaction between Feyerabend and Popper for what concerns their work on quantum foundations. However, as recalled in the introduction, it would be impossible to understand these developments without referring to their peculiar personal relationship. As we shall see in this section, not only Popper was an intimate confidant of Feyerabend on very personal matters (on love relationships, sexual problems, psychological uncertainties, etc.), but almost a father figure to Feyerabend, showing towards his pupil a mixture of unconditional support and authoritative lessons of discipline. Feyerabend, on the other hand, acted on several occasions as a rebellious child, sometimes even lying to Popper to get out of trouble, but eventually always coming back to ask for guidance and support when needed. This until he probably realized that to reach full independence he needed to completely cut his relationship with Popper. I reconstruct here the development of a less academic and more personal relationship between Popper and Feyerabend and show how relevant this aspect has been for the departure of Feyerabend from Popper's positions, firstly on the FQM and then in the general philosophy of science.

It has been already recalled that Popper had been the single main influence on Feyerabend's formative years, as the latter stated in one of his early CVs in 1951: "radical change to physics and gradual approach to philosophy, mainly under the influence of Prof. K. R. Popper […]". (quoted in Collodel 2016). Later on, Feyerabend kept acknowledging Popper's pivotal influence in his lectures on the philosophy of science, for instance when in Bristol: "Lectures + tutorials go on very well indeed. It seems that people find my lectures on philosophy of science interesting, at least some told me so and the discussions are very lively. But I know that I would not have given such lectures without having heard your lectures" (Feyerabend to Popper 14.10.1955 N.30; in Collodel and Oberheim 2020).

Perhaps more remarkable, however, is the high opinion that Popper had towards this young student. Popper's appreciative words give us a hint of how much he valued Feyerabend as a pupil when the latter turned down the offer to become his research assistant:

> I was very much looking forward to having you here, and I had a lot of interesting work for you. […] If […] you still feel like coming, you will be most welcome to me. So, please, do not think that I have changed my mind. On the contrary, the prospect of loosing you has made it clear to me how much I would lose. (Popper to Feyerabend 02.01.1954 N. 1; in Collodel and Oberheim 2020).

And throughout the whole decade of 1950s, and more seldom in the early 1960s, Feyerabend also asked literally tens of letters of recommendations, and career advice, which Popper always granted.

Moreover, their relationship largely transcended the normal interaction between a pupil and his mentor. In fact, Feyerabend used to tell Popper intimate details of his own private life and ask pieces of advice on very personal matters. For instance, he used to ask Popper advice on, for instance, whether he should tell his girlfriend about his own sexual problems (his war injuries had left him impotent), or whether he should propose, or he once told Popper about an affair he had. Just to provide one example, among the many ones, of the depth that their personal relationship had reached by 1956, it is interesting to quote a passage from a letter that Feyerabend wrote to Popper when the former started dating his second wife to come, Mary O'Neill:

> I told her practically everything about myself, about my past, my way of acting, my difficulties (of body and mind) and this was the first time that I was completely open to somebody from the very beginning. Pushed by circumstances, certainly, for I do not know whether I would have acted like this without your insistence. I told her about your suggestion and about the reasons you gave and I asked her not to try to see me for 14 days and consider everything carefully. […] [H]aving told her everything, having had this kind of response, I feel more attached to her as ever. I was afraid to tell you this and this is the reason for my silence. Now, that I am writing this I feel in a

> curious way excited and am not far from crying. But – all that does not mean
> that I am not always trying make [sic] the decision you want me to do.
> (11.02.1956 N. 1; in Collodel and Oberheim 2020).[10]

After the period spent together in Berkeley in 1962, Popper and Feyerabend's correspondence became sparser. However, one can still find evidence of the deep consideration that Feyerabend used to have for Popper: "I am looking forward to every new paper from you. […] [R]eading you makes me optimistic again and convinces [*sic*] that it is worthwhile to philosophize." (Feyerabend to Popper 19.12.1963 N. 3; in Collodel and Oberheim 2020).

However, a curtain of clouds –which actually appeared quite early after they started collaborating– covered the brightness of their relationship on several occasions. A first episode of discordance between Popper and Feyerabend happened already after Feyerabend published one of his first papers (Feyerabend 1954b). Popper was unhappy with the way that Feyerabend cited him because his ideas were not properly acknowledged (see Collodel 2016). To such a criticism Feyerabend replied:

> [T]here is a peculiar psychological phenomenon. I can hardly underrate what I have learned from you both in content and in the attitude toward philosophical and even scientific problems. But […] I am discovering new aspects of your ideas and possibilities of application. [*sic*] Then *I am sometimes thinking that it was my invention* (emphasis are mine; quoted in Collodel 2016).

We will see that this has always been the line of defense that Feyerabend adopted against the various allegations of plagiarism by Popper.

A similar, but much more serious incident happened only a couple of years later. As recalled in section 2, in April 1956, Feyerabend submitted a short note (Feyerabend 1956) with his criticism of von Neumann's impossibility proof to the prestigious German physics journal *Zeitschrift für Physik*. Popper accused Feyerabend of having misappropriated his and Joseph Agassi's –Popper's assistant at that time– ideas:

> I have read your new paper. I do agree of course with its contents, since you got the complete contents of this paper here, in my room, from Joske [Joseph Agassi] and myself. […] I may remind you again of the fact that, when you still believed in the Neumann Proof you were here, and we had a great fight about it. […] In fact, we made the point much clearer and better than you do it now in your paper. […] If you do not add a very full acknowledgement to Joske and myself, it will be a flagrant piece of plagiarism. I have warned you before against this practice. I think it would be quite wrong if I would close my eyes to this case. I do not intend to do this, and I am now warning you

---

[10] Agassi recalled in various occasions (see Agassi 1993) that Popper interfered a great deal with Feyerabend's private life.

again. (Popper to Feyerabend 16.04.1956 N.17; in Collodel and Oberheim 2020).

And a few days later again:

> I do not believe that the contents of your paper are really worth publishing. I should have never published it myself. It is thus not that I think you are taking valuable goods from Joske and myself – goods which we find valuable. It is a different thing: a matter of principle. The goods are goods which you find valuable enough to publish. And they are not your property. […] I shall express my view that I consider a publication to be morally wrong. […] I should have willingly agreed to have it published by you, provided you would have proceeded in the proper way. […] I have to do it for your sake: otherwise you will never learn what one may do and not do (Popper to Feyerabend 18.04.1956 N.18; in Collodel and Oberheim 2020).

Under Popper's insistence, Feyerabend accepted to include a note written by Popper, acknowledging his and Agassi's contribution. However, he "eventually prevented the note from appearing, confessing to an inquiring Agassi that he thought that the acknowledgement which Karl [Popper] had dictated was much too strong, but [he] ha[d] been afraid of quarrel" (Collodel 2016). Yet, at the same time, Feyerabend falsely declared to Popper: "Although I had sent the footnotes suggested by you on the very same day I told you, the proofs (which arrived much later) did not contain them and I had to add them from my memory, as I had already lost the original MS." (Feyerabend to Popper undated 1956 N.22; in Collodel and Oberheim 2020).

Even before that, on more than one occasion, Feyerabend had expressed sorrow for his general behaviors towards Popper: "I am very sorry for all this and I hope that this time where I only cause disappointment to you will soon end." (Feyerabend to Popper 01.01.1956 N. 1; in Collodel and Oberheim 2020), but also showed an uneasiness to comply with Popper's demands: "sometimes I feel inclined to act as you want because you want it and because I do not want to disappoint you." (Feyerabend to Popper, undated 1956 N. 4; in Collodel and Oberheim 2020).

Popper, for his part, never missed the occasion to patronizing Feyerabend. When Feyerabend gave his first lecture course on FQM (see section 2), Popper accused him of overstating his results, to which Feyerabend submissively replied: "I think you are completely right. I should be more modest in my lectures; […] I shall change my attitude and procedure." (Feyerabend to Popper 20.02.1956 N. 1; in Collodel and Oberheim 2020).

In the early 1960s, the allegations of misappropriation of Popper's ideas by part of Feyerabend became graver. Two of the most distinguished and loyal among Popper's pupils, Imre Lakatos and William Bartley III, started reporting on Feyerabend misbehaviors towards Popper. Lakatos studied in detail a series of papers by Feyerabend and annotated all the passages where Popper's ideas were unduly used against him or cited without any acknowledgement, whereas Bartley –who spent the years 1962-63 in Berkeley– regularly reported to Popper, maintaining that "in his university courses Feyerabend was basically giving Popper's LSE lectures, yet without ever mentioning Popper's name." (quoted in Collodel 2016).

In 1964, Popper wrote to Hans Albert –a German philosopher follower of the Popperian school– asking for his advice on how to deal with Feyerabend plagiarism:

> As you well know, I am an old friend and admirer of Paul Feyerabend and as you are also a friend of his, I would like to ask your advice about a rather awkward situation. […] I have some letters from Paul Feyerabend in which he speaks very highly of me. […] You have to admit that this is very nice; and I am sure that it was meant sincerely. […] Unfortunately our friendship has been somewhat clouded by the fact that he is neurotic and that his neurosis partly consists in him stealing my ideas like a raven for many years […] He mentions me somewhere in the respective articles, sometimes even quite frequently; but not in the parts about his "own" contributions, which are usually stolen from me. However, his "own" contributions are often defended against me […].
>
> Well, I am used to that. I do not take it too hard. After all I have enough ideas and may leave some of them for my students (though without being asked), even if perhaps it goes a bit too far that my ideas are (a) stolen from me, and (b) used to attack me. […] These remarks […] are a kind of "Stop, thief!" Because poor Paul knows he is stealing: I often called his attention to this in a friendly way. The last time (in March in Berkeley, 1962) he answered: "*Your ideas are so original that it takes a great effort to assimilate them; and by the time one has assimilated them one thinks they are one's own.*" (original emphasis; Popper to Hans Albert on 17.02.1964 Reproduced in English translation in note 4 to the letter: Feyerabend to Popper 27.12.1965 N. 4; in Collodel and Oberheim 2020).

And yet the accusation of misappropriation continued. On the occasion of a *Festschrift* for Herbert Feigl –a logical empiricist and former member of the Vienna Circle that was a major influence on Feyerabend when he moved to the US– Feyerabend wrote a paper (Feyerabend 1966) which he sent to Popper for feedback. Once more, the latter accused Feyerabend of having made use of his ideas without proper acknowledgement and intimated him to add a footnote –which Feyerabend agreed to add and then did not, again telling Popper that it was too late. Feyerabend wrote to Popper adducing a general justification to his behavior –the same he had already used a decade before:

> [I]f there is another occasion where you think that I have made use of your ideas without acknowledging them, please let me know, […] so that the matter can be cleared up that is that I can either explain to you or realise myself that I had forgotten something. For I do not think that I ever consciously used your ideas without the proper acknowledgement although it may have been the case that having heard them from you I continued thinking,

> returning in a roundabout manner to the ideas I had originally gotten from you, but without knowing any longer that this was the case (this, I think, is what happened in the case of my short note on the Neumann proof)." (Feyerabend to Popper 27.12.1965 N. 4; in Collodel and Oberheim 2020)

Finally, coming back to the harsh criticism of (Popper 1967), Feyerabend confessed to John W. Watkins –another prominent exponent of the Popperian school– that behind his paper there were not only scientific reasons but personal resentment too:

> one of the reasons why I was mad at Popper was that his paper [(Popper, 1967)] did not pay any attention to my criticism of 1962 [(Feyerabend 1962)]. Maybe he had not read my paper (which I sent him); maybe he did not like it." (Feyerabend to Watkins on 17.12.1967, quoted in Collodel 2016).

The relationship between Popper and Feyerabend was at that point irremediably undermined. Popper did not even read Feyerabend papers against his views on FQM (Feyerabend 1968; Feyerabend 1969). In answering to Landé's request for a rebuttal (see section 2), Popper showed with a certain resignation that he had given up fighting with Feyerabend:

> <u>I heard about it</u>. I do not want to read it because why should I get angry? (Feyerabend is of course one of my former students for whom I did more than any teacher can be expected to do and has behaved to me in return simply disgustingly.) I should be most grateful for your defense if you would defend me. [...] I do not wish to defend myself because this would mean reading this silly stuff. (Popper to Landé; published in Del Santo 2019a)

He also replied to Margenau, offering an explanation for Feyerabend's ingratitude, which leaves us with a somewhat sour taste, for the end of a relationship that, through thick or thin, has changed the conception of philosophy of science in the last century:

> Feyerabend was once a student of mine, and I treated him extremely well. He never got over it: *there are people who can never forget a benefit received, and other who cannot forgive it*. Moreover, Feyerabend has no original ideas, but he poses as an original thinker: he is a compulsive plagiarist. He has stolen many of my ideas and although sometimes admitted this, he continues to criticise me, using my own ideas in his criticisms (emphasis are mine; Popper to Margenau on 21.02.1969; published in Del Santo 2019a) .

### 4. Conclusions

John Preston has commented that "there is something obsessive about the way in which Popper became Feyerabend's favourite whipping-post" (Preston 1997). In this paper, I have reconstructed how the rift between Popper and Feyerabend came about, for what concerns the main topic of their discussions for about two decades: the foundations of quantum mechanics. I have also shown that the personal component of a peculiarly strong relationship between them played a substantial role in this process. Indeed, indulging in some amateur psychology, it seems that Feyerabend –which was treated by Popper as his (academic) "son"– "disliked Popper's growing tendency to preach authoritatively" (Oberheim 2006; taken from Collodel 2016) and deeply suffered from his "unwanted indebtedness to Popper" (Watkins 2000). This led him to assume more and more the role of a rebellious "son", and when the time was ripe he entered his (academic) "adolescence", in which he violently rejected his "father's" idea(l)s at any cost.

Perhaps philosophy of science would have lost something without such a conflict.

**Acknowledgments**

I acknowledge the financial support through a DOC Fellowship of the Austrian Academy of Sciences (ÖAW).